%% file: article.tex
\documentclass[review]{elsarticle}

\usepackage{hyperref}
\usepackage{lineno}
\modulolinenumbers[0]

\usepackage{amsmath,amssymb}
\usepackage{siunitx}
\usepackage[T1]{fontenc}
\usepackage[utf8]{inputenc}

\journal{Physics Letters B}

\newcommand{\diff}{\,\mathrm{d}}
\newcommand{\e}{\mathrm{e}}
\newcommand{\Ei}{\operatorname{Ei}}
\newcommand{\figref}[1]{Fig.~\ref{#1}}
\newenvironment{Align}{\linenomath\align}{\endalign\endlinenomath}
\newenvironment{Equation}{\begin{linenomath}\begin{equation}}{\end{equation}\end{linenomath}}
\begin{document}

\begin{frontmatter}

\title{Radiative corrections to the average bremsstrahlung energy loss of
 high-energy muons}
\author[TUDo]{A. Sandrock}
\author[MEPhI]{S. R. Kelner}
\author[TUDo]{W. Rhode}
\address[TUDo]{Department of Physics, TU Dortmund University, D-44221 Dortmund,
 Germany}
\address[MEPhI]{National Research Nuclear University (MEPhI),
 Kashirskoe Shosse 31, Moscow 115409, Russia}

\begin{abstract}
High-energy muons can travel large thicknesses of matter. For underground neutrino
and cosmic ray detectors the energy loss of muons has to be known accurately for
simulations. In this article the next-to-leading order correction to the average
energy loss of muons through bremsstrahlung is calculated using a modified
Weizsäcker-Williams method. An analytical parametrisation of the
numerical results is given.
\end{abstract}


\end{frontmatter}


\input{1-introduction.tex}
\input{2-method.tex}
\input{3-results.tex}
\input{4-conclusion.tex}
\input{acknowledgement.tex}
\section*{References}
\bibliography{references.bib}
\end{document}

%% file: 1-introduction.tex
\section{Introduction}
The muon bremsstrahlung cross section has been studied extensively for many
years \cite{PetrukhinShestakov,ABB1,ABB2,KKP,KKP-atomic}. Together with the
production of electron-positron pairs \cite{KelnerKotov,KokoulinPetrukhin1,%
KokoulinPetrukhin2,Kelner-atomic} and the inelastic nuclear interaction
\cite{ALLM91,ALLM97,BugaevShlepin} it describes the dominant contribution to
the energy loss of high-energy muons.

Muons with energies of tens to hundreds of TeV can travel distances of the
order of several kilometers. Therefore it is necessary to know the average
energy loss per unit length
\begin{linenomath*}
\begin{equation}
-\left\langle\frac{\diff E}{\diff x}\right\rangle 
 = N \int E v \frac{\diff\sigma}{\diff v} \diff v
\end{equation}
\end{linenomath*}
accurately. Here $v = (E - E')/E$ is the relative energy loss per interaction,
and $N$ is the number density of target atoms. Previous calculations took into
account the modification of the Coulomb interaction with the nucleus by elastic
and inelastic nuclear form factors, the contribution of atomic electrons as
target for muon bremsstrahlung and the inelastic interaction with the target
nucleus. This article discusses the correction of the energy loss through
virtual and real radiative corrections. Since this correction is small compared
to the main contribution, we restrict our treatment of the nucleus to elastic
atomic and nuclear form factors.

The energy loss is of importance for underground detectors for two reasons:
on the one hand, the energy loss is needed to predict the spectrum of muons
that will reach the detector; on the other hand the energy lost by a muon
inside the detector on a given length is used to reconstruct the energy of the
radiating particle. The energy reconstruction is further complicated by 
its sensitivity to the distribution of energy losses and their correlation to
the energy of the muon. Especially rare large stochastic energy losses enlarge
the variance of the energy loss per unit length. As a first step to revisit
this problem, in the present article, the muon energy dependent average energy
loss per length is calculated.

In the calculation of radiative corrections in QED processes with virtual
photons give rise to logarithmically divergent integrals; to obtain a finite
result, it is necessary to add the cross section for the emission of an
additional photon with energy $\omega < \omega_\text{min}$ which cancels this
divergence. Usually $\omega_\text{min}$ is identified with the finite energy
resolution of the detector and assumed to be small compared to the mass of the
radiating particle, such that the approximation of classical currents can be
used. The contribution of harder photons indistinguishable from a single photon
is then evaluated numerically according to the conditions of the experiment (see
e.~g. \cite{Arbuzov}). In the problem of muon propagation, however, the particle
may traverse several kilometers of material before the energy losses can be seen
by the detector. Therefore the cross section has to be integrated over all
kinematically allowed states of the additional photon. So the energy loss
depends only on the primary energy of the muon.

Unless stated otherwise, all equations are presented in a system of units where
$\hbar = c = m_\mu = 1$.

%% file: 2-method.tex
\section{Method}
The calculation reported here is based on the Weizsäcker-Williams method
\cite{Weizsaecker,Williams}, which approximates the effect of a nucleus by a
spectrum of equivalent photons. This method allows to express the
bremsstrahlung cross section through the Compton cross section convolved with
the equivalent photon flux. Using the radiative corrections to the Compton
effect in \cite{BrownFeynman}, the radiative corrections to the bremsstrahlung
spectrum were first calculated in the soft-photon approximation in \cite{%
MorkOlsen} for an unscreened or totally screened nucleus.

\subsection{Conventional Weizsäcker-Williams method}
\begin{figure}
\centering{
\includegraphics[width=0.7\textwidth,angle=0]{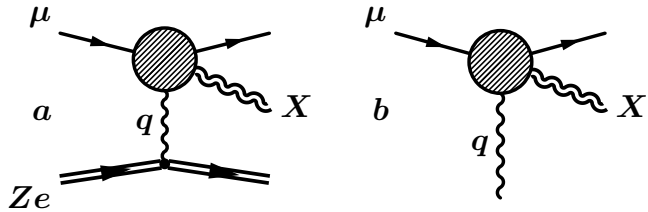}
}
\caption{$(a)$ Symbolic Diagram of interaction of a muon with an atom.
$(b)$ Equivalent Diagram in the Weizsäcker-Williams method. The shaded blob
denotes the internal part of the diagram. $X$ denotes particles created in the
collision.}
\label{fig:WW_dia}
\end{figure}

Considering the collision of a fast muon with an atom, we introduce two systems
of reference: the laboratory system $K_Z$ in which the atom is at rest and the
muon has a Lorentz factor $\gamma \gg 1$, and the system $K_\mu$ in which the
muon is at rest and the atom has a Lorentz factor of $\gamma$. The interaction
of the muon with the atom can be described symbolically by the diagram in
\figref{fig:WW_dia}$(a)$. The shaded blob denotes the internal part of the
diagram, the double line $X$ denotes particles created in the collision.
Diagram \figref{fig:WW_dia}$(b)$ describes a similar process due to collision
of a real photon with a muon. We consider two cases here:
\begin{enumerate}
\item $X = \gamma$. In this case $(a)$ is bremsstrahlung and we assume that the
blob also contains radiative corrections; $(b)$ is Compton scattering with
radiative corrections.
\item $X = 2\gamma$. In this case $(a)$ is double bremsstrahlung and $(b)$ is
double Compton scattering.
\end{enumerate}

In the Weizsäcker-Williams method the field of the atom in $K_\mu$ is replaced
by a flux of equivalent photons with the spectrum $n(\omega) \diff\omega$. This
allows to relate the differential cross sections of the processes $(a)$ and
$(b)$ by the relation
\begin{Equation}
\diff\sigma_a = \diff\sigma_b n(\omega) \diff\omega,
\label{eq:WW}
\end{Equation}
and for the total cross section
\begin{Equation}
\int\diff\sigma_a = \int\diff\sigma_b n(\omega) \diff\omega.
\end{Equation}
It is convenient to calculate the cross section $\sigma_b$ in the frame $K_\mu$.
Since the cross section is Lorentz-invariant, the transition from $K_\mu$ to
$K_Z$ is trivial.

In $K_Z$ the average energy loss per unit length caused by the process $(a)$ is
given by
\begin{Equation}
-\left\langle\frac{\diff E}{\diff x}\right\rangle = N E \Sigma,\quad
\Sigma = \frac{1}{E} \int (E - E') \diff\sigma_a.
\end{Equation}
where $E$ ($E'$) is the initial (final) muon energy, $N$ is the number density
of target atoms per unit volume. The quantity $\Sigma$ can be rewritten in a
relativistically invariant form as
\begin{Equation}
\Sigma = \frac{1}{(up)} \int ((up) - (up')) \diff\sigma_a,
\end{Equation}
where $u$ is the 4-velocity of the atom and $p, p'$ are the initial and final
4-momenta of the muon respectively, $(up) = u^0 p^0 - \mathbf{up}$ is the
scalar product of 4-vectors. Using \eqref{eq:WW} it is possible to rewrite this
as
\begin{Equation}
\Sigma = \int \frac{(up) - (up')}{(up)} \diff\sigma_b n(\omega) \diff\omega.
\end{Equation}
In this equation we will calculate the integrand in the $K_\mu$ frame. For
Compton scattering the energy-momentum
conservation gives $(u,p - p') = (u, q' - q)$, where $q$ ($q'$) is the initial
(final) photon 4-momentum. In the frame $K_\mu$ we have $(uq)/\gamma = \omega
(1 - \beta) \approx \omega/(2\gamma^2)$, where $\omega$ is the initial photon
energy. Since $\omega \sim \gamma$, the ratio $(uq)/\gamma \sim 1/\gamma$ is
negligible. The other term is
\begin{Equation}
\frac{(uq')}{\gamma} = \omega'(1 - \beta \cos \theta)
  \approx \frac{1}{2} \omega' (\theta^2 + 1/\gamma^2).
\end{Equation}
In Compton scattering $\theta \lesssim 1/\sqrt{\gamma}$
\cite{BeresteckiLifshicPitaevski} and the second term in parentheses is
negligible. Therefore we obtain for the first case
\begin{Equation}
1 - \frac{(up')}{(up)} \approx 1 - \frac{\omega'}{\omega}.
\end{Equation}
Similarly for double Compton scattering we have
\begin{Equation}
1 - \frac{(up')}{(up)} \approx \omega_1 (1 - \cos\theta_1)
 + \omega_2 (1 - \cos\theta_2),
\label{eq:double}
\end{Equation}
where $\omega_{1,2}, \theta_{1,2}$ are the energies and angles of the final
photons.

\subsection{Modified Weizsäcker-Williams method}
For a point-like nucleus, the pseudophoton flux in the rest frame of the muon is
given by \cite{BeresteckiLifshicPitaevski}
\begin{Equation}
n(\omega) \diff\omega = \frac{2}{\pi} Z^2 \alpha \frac{\diff\omega}{\omega}
  \ln \frac{\gamma}{\omega}
\end{Equation}
where $\gamma$ is the Lorentz factor of the incident particle in the laboratory
frame, or, equivalently the Lorentz factor of the nucleus in the rest frame of
the muon.

For muons, it is necessary to take into account the extended nucleus and the
screening of the nucleus by atomic electrons, because the characteristic
momentum transfer of $q \sim m_\mu$ is comparable to the inverse radius of the
nucleus and the minimum momentum transfer $\delta \sim m_\mu^2/E$ is comparable
to or smaller than the inverse radius of the atom \cite{ABB2}.

For an atom with nuclear and atomic formfactors $F_\text{n}(q^2),
F_\text{a}(q^2)$ we obtain
\begin{Equation}
n(\omega)\diff\omega=\frac{\alpha Z^2\diff\omega}{\pi \omega}
\int\limits_{\omega^2/\gamma^2}^\infty
\frac{\tau - \omega^2/\gamma^2}{\tau^2}\,
\left( F_\text{n}(\tau) - F_\text{a}(\tau)\right)^2 \diff\tau.
\label{eq:logfactor}
\end{Equation}

In this work the charge distribution of the nucleus and of the atomic
electrons are described by a Gaussian and an exponential distribution,
respectively, resulting in the form factors
\begin{Align}
F_\text{n}(q^2) &= \exp\left[-\frac{q^2 R_\text{n}^2}{6}\right],\label{Fn}\\
F_\text{a}(q^2) &= \left[1 + \frac{q^2 R_\text{a}^2}{12}\right]^{-2}\label{Fa}
\end{Align}
with $R$ the Rms-radius of the charge distribution. The atomic and nuclear
radius can be parametrised for light and medium nuclei as
 \cite{ButkevichKokoulin}
\begin{Align}
R_\text{n} &= 1.27 A^{0.27} \si{fm},\\
R_\text{a} &= \frac{183 Z^{-1/3}}{2.718 m_\text{e}}
\intertext{with $Z$ the nucleus charge and $A$ its mass number, and for
hydrogen}
R_\text{p} &= \SI{0.85}{fm}.
\end{Align}

With the above formfactors, the calculation of the pseudophoton flux integral
gives
\begin{Align}
n(\omega)\diff\omega = \frac{\alpha Z^2}{\pi} \left[\frac {1}{(bz+1)^2}
  \left( -\frac {17}{6} - 4 b^2 z^2 - 7bz \right)\right.\nonumber\\
- 2\exp\left(\frac {a}{b}\right) ( az - 2 bz + a/b-1) \Ei(1, a(z + 1/b))
\label{num1}\\
  - (2 az + 4 bz + 2) \Ei(1, az) + (2 az + 1) \Ei(1, 2 az)\nonumber\\ 
\left. \phantom{\frac11}+ (4 bz + 1) \ln  \left(1 + \frac{1}{bz}
  \right) -\e^{-2 az} + 4\e^{-az} \right] 
 \frac{\diff\omega}{\omega},\nonumber
\end{Align}
where $\Ei(1, x) = \int_x^\infty e^{-t}/t \diff t$ is 
the exponential integral, $z = \omega^2/\gamma^2$, $a = R_\text{n}^2/6$,
$b = R_\text{a}^2/12$.

\subsection{Corrections to energy loss}
The contribution to the  energy loss of a pseudophoton with energy $\omega$ in
the rest system of the charged projectile can be conveniently expressed by the
angles and energies of the final particles in this system as
\begin{Equation}
\gamma \int \left(1 - \frac{\omega'}{\omega}\right)
 d\sigma(\omega, \omega')
\end{Equation}
for virtual radiative corrections, where $d\sigma(\omega, \omega')$ is the
differential cross section for Compton scattering and the scattering angle
$\theta$ is determined by conservation laws; and
\begin{Equation}
\gamma \int [\omega_1(1 - \cos \theta_1) + \omega_2(1 - \cos \theta_2)]
 d\sigma(\omega, \omega_1, \omega_2, \theta_1, \theta_2)
\end{Equation}
for the double bremsstrahlung contribution, where $d\sigma(\omega, \omega_1,
\omega_2, \theta_1, \theta_2)$ is the differential cross section for double
Compton scattering, dependent on the initial and final photon energies $\omega$
and $\omega_1, \omega_2$ and the scattering angles $\theta_1, \theta_2$; the
azimuthal angle between the two photons is determined by conservation laws.
All quantities refer to the rest system of the muon. This formulation
improves the numerical stability compared to using the angles in the lab frame,
because the cross section is strongly peaked in the forward direction, while
this peak is much broader in the rest frame.

In addition, the contribution from vacuum polarization was calculated directly.
The loop correction to the virtual photon coupled to the atom can be
interpreted as a factor modifying the form factor of the atom.

%% file: 3-results.tex
\section{Results}
\begin{figure}
\centering{
\includegraphics[angle=-90,width=0.7\textwidth]{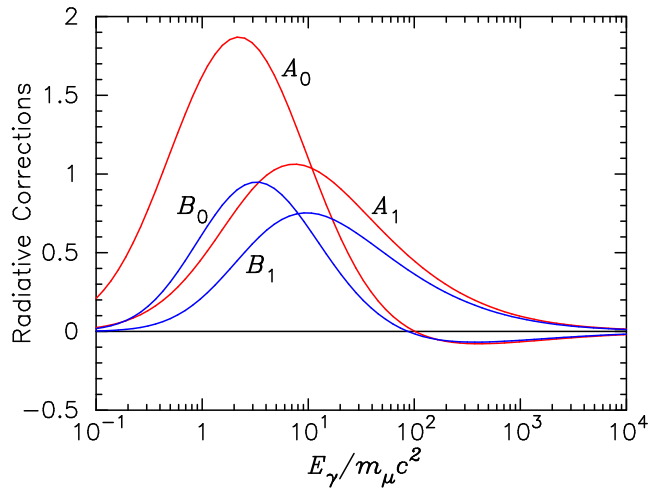}
}
\caption{Numerical results of the integration over angles and energies for the
contribution of the pseudophoton with energy $E_\gamma$ for the functions
defined in \eqref{eq:rad_corr1}, \eqref{eq:rad_corr2}.}
\label{fig:rad_corr}
\end{figure}

Let us represent Eq.~(30) from \cite{BrownFeynman} in the form
\begin{Equation}
P = P_0 + P_1 \ln \lambda
\end{Equation}
where $\lambda$ is the fictitious photon mass necessary to regularize the
infrared divergences which cancels out in the final result. The functions $P_0$
and $P_1$ depend only on the initial and final photon energies $\omega,
\omega'$. Then the virtual radiative corrections to the differential Compton
cross section are given by
\begin{Equation}
\diff\sigma_\text{corr} = -\alpha r_\mu^2 (P_0 + P_1 \ln \lambda)
 \frac{\diff\omega}{\omega^2}
\end{Equation}
and we obtain
\begin{Equation}
\sigma_\text{corr} = -\frac{\alpha r_\mu^2}{\omega^2} \int\limits_{\omega_*}%
^\omega (P_0 + P_1 \ln \lambda) \diff\omega'
\end{Equation}
for the total cross section, where $\omega_* = \omega/(1 + 2\omega)$. To
determine the correction to the average energy loss we have to calculate the
integral
\begin{Equation}
-\frac{1}{\omega^2} \int\limits_{\omega_*}^\omega (P_0 + P_1 \ln \lambda)
 \left(1 - \frac{\omega'}{\omega}\right) \diff\omega'.
\end{Equation}
The results can be expressed through the following functions:
\begin{Align}
A_i &= -\frac{1}{\omega^2} \int\limits_{\omega_*}^\omega P_i \diff\omega'
  \label{eq:rad_corr1}\\
B_i &= -\frac{1}{\omega^2} \int\limits_{\omega_*}^\omega P_i
  \left(1 - \frac{\omega'}{\omega}\right) \diff\omega',
  \label{eq:rad_corr2}\\
i &= 1, 2 \nonumber.
\end{Align}
The graphs of the functions $A_{0,1}, B_{0,1}$ which only depend on $\omega$
are shown in \figref{fig:rad_corr}.

\begin{figure}
\centering{
\includegraphics[angle=-90,width=0.7\textwidth]{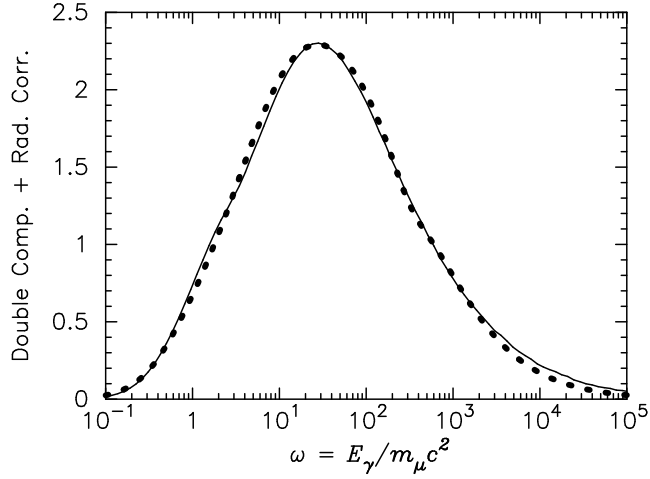}
}
\caption{Comparison of the numerical results (solid line) and the analytical
parametrisation \eqref{eq:double_para} (dots).}
\label{fig:double}
\end{figure}

For the calculation of the integral \eqref{eq:double} we used the cross section
from \cite{Skyrme}. The integral diverges as $\omega_{1,2} \rightarrow 0$, so
we integrated over the region $\omega_{1,2} > \omega_\text{min}$. Let us
denote the result of this integration as $\alpha r_\mu^2 \gamma X(\omega,
\omega_\text{min})$. The sum
\begin{Equation}
f(\omega) \equiv X(\omega, \omega_\text{min}) + B_0(\omega) + B_1(\omega)
  \ln (2\omega_\text{min})
\end{Equation}
in the limit $\omega_\text{min} \rightarrow 0$ does not depend on
$\omega_\text{min}$. Numerical calculations for $\omega_\text{min} = 10^{-4}$
and $\omega_\text{min} = 10^{-5}$ give practically the same result (here we use
the known relation $\lambda = 2 \omega_\text{min}$). For the convenience of
further calculations we have obtained the following approximate formula:
\begin{Equation}
f(\omega) = 405 \omega
  \left[1 - \frac{(0.006 \omega)^2}{16 + (0.006 \omega)^4} \right]
  \frac{\ln(1 + 0.00654 \omega)}{1 + 4 \ln^2(\omega + 1) + \omega^2},
\label{eq:double_para}
\end{Equation}
The comparison between the numerical results and this approximation is shown in
\figref{fig:double}.

By numerical integration over the product of the pseudophoton flux and this
function the energy loss is obtained. It can be parametrized by the function
\begin{Equation}
-\frac{1}{E} \left\langle\frac{\diff E}{\diff x}\right\rangle = Z^2 \alpha^2 N
  r_\mu^2 c_1 \ln \frac{R_\text{a}/(R_\text{n} c_2)}
  {1 + R_\text{a}/(R_\text{n} c_2) \cdot c_3/\gamma}
\end{Equation}
with
\begin{Align}
c_1 &= \num{8.303},\\
c_2 &= \num{0.999},\\
c_3 &= \num{4.099} + \num{6.335} Z^{1/3}.
\end{Align}

The additional contribution of vacuum polarization can be parametrized as
\begin{Equation}
-\frac{1}{E} \left\langle \frac{\diff E}{\diff x} \right\rangle = Z^2 \alpha^2 N
  r_\mu^2 3 \ln \frac{Y}{1 + \xi Y/(\gamma R_\text{N})}
\end{Equation}
with
\begin{Align}
Y &= \frac{R_\text{a}}{16.7 R_\text{N}},\\
\xi &= 0.1 + 0.0921 A^{0.8}.
\end{Align}

\subsection{Ratio to the main contribution}
When the main contribution to the bremsstrahlung energy loss is calculated in
the modified Weizsäcker-Williams method with the above form factors, the ratio
between the radiative correction and the main contribution is independent of
$Z$ and $A$ and can be parametrized as 
\begin{Equation}
-\frac{1}{E} \left\langle \frac{\diff E}{\diff x} \right\rangle_\text{rad} = 
  -\frac{1}{E} \left\langle\frac{\diff E}{\diff x}\right\rangle_0 \alpha
  \delta(\gamma)
\end{Equation}
with
\begin{Equation}
\delta(\gamma) = 2.66 \frac{\xi - 1}{\xi + 1}, \quad
  \xi = \num{0.0406} \sqrt{\gamma} + 3.39
\end{Equation}
for light and medium nuclei and
\begin{Equation}
\delta(\gamma) = 2.77 \frac{\xi - 1}{\xi + 1}, \quad
   \xi = 0.02 \sqrt{\gamma} + 5.32.
\end{Equation}
for hydrogen.

%% file: 4-conclusion.tex
\section{Conclusion}
The radiative corrections to the average energy loss of muons by bremsstrahlung
have been calculated using a modified Weizsäcker-Williams method. It was found
that the energy loss is increased by about \SI{2}{\%} in the complete screening
regime. An analytical parametrization of the numerical results has been
obtained.

\begin{figure}
\centering{
\includegraphics[width=0.5\textwidth,angle=-90]{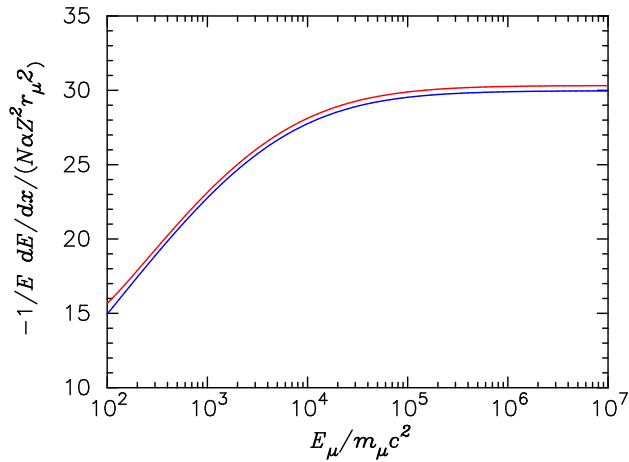}
}
\caption{Energy losses in oxygen due to bremstrahlung. The red line shows the
calculation using the Weizsäcker-Williams method with the photon spectrum
\eqref{num1}, the blue line is the usual calculation with the formfactors
\eqref{Fn}, \eqref{Fa}. The difference between the curves is approximately
\SI{1}{\%}.}
\label{fig:ww_exact}
\end{figure}

To estimate the accuracy of the calculation the contribution of leading order
bremsstrahlung was calculated using an exact calculation and the Weizsäcker-%
Williams method with the used formfactors. The difference between the exact
numerical results and the result by the Weizsäcker-Williams method is
approximately \SI{1}{\%} (cf. Fig.~\ref{fig:ww_exact}).

This result is important for underground muon and neutrino detectors,
because it is a step to reduce the uncertainty on the energy loss of muons on
their passage to the detector.

%% file: acknowledgement.tex
\section*{Acknowledgements}
A.~S. and W.~R. acknowledge funding by the Helmholtz Allianz für
Astroteilchenphysik. A.~S. gratefully acknowledges the hospitality of MEPhI
during this work. The authors thank A.~A.~Petrukhin and R.~P.~Kokoulin for
valuable discussions.